# Temporal Rainbow Scattering at Boundary-Induced Time Interfaces

L. Stefanini[†], D. Ramaccia, A. Toscano, F. Bilotti

Department of Industrial, Electronic and Mechanical Engineering ROMA TRE University, Via Vito Volterra 62, 00146 Roma, Italia

## ABSTRACT

Since the dawn of modern optics and electromagnetics, optical prism is one of the most fascinating optical elements for refracting light. Exploiting its frequency dispersive behaviour, a prism is able to refract different frequencies in different directions, realizing polychromatic light rainbows. Recently, thanks to their engineerable electromagnetic response, metamaterials have been exploited for achieving novel refractive scattering processes, going beyond the classical prism effects. In this Manuscript, we report on a novel rainbow-like scattering process taking place at the interface of a boundary-induced temporal metamaterial realized by instantaneously opening the boundary conditions of a parallel plate waveguide. Changing abruptly the conductivity of one of the two metallic plates, we demonstrate that an equivalent temporal interface between two different media is realized, and the monochromatic wave propagating into the waveguide gets scattered into a polychromatic rainbow in free-space. We derive the relationships between the waveguide mode and the raising rainbow in terms of scattered amplitude and frequencies as a function of the elevation angle with respect to the waveguide axis. We apply the underlying physics to control the temporal rainbow by imposing a principal direction of scattering by design. Full-wave numerical simulations are performed for computing the rainbow temporal scattering and verifying the design guidelines for achieving controlled temporal rainbow scattering

## 1. Introduction

Rainbows are optical phenomena that take place when the frequency components of a polychromatic light get scattered into different directions. A classic example is reported by Newton [1], who observed the separation of white light into its fundamental colors passing through a prism made of glass. This behavior is justified by the frequency dispersive optical properties of glass: each frequency component observes a different refractive index and gets scattered in a different direction according to the Snell's law at the air-glass interface. The material properties of the prism play a fundamental role in the design and control of the rainbow scattering process.

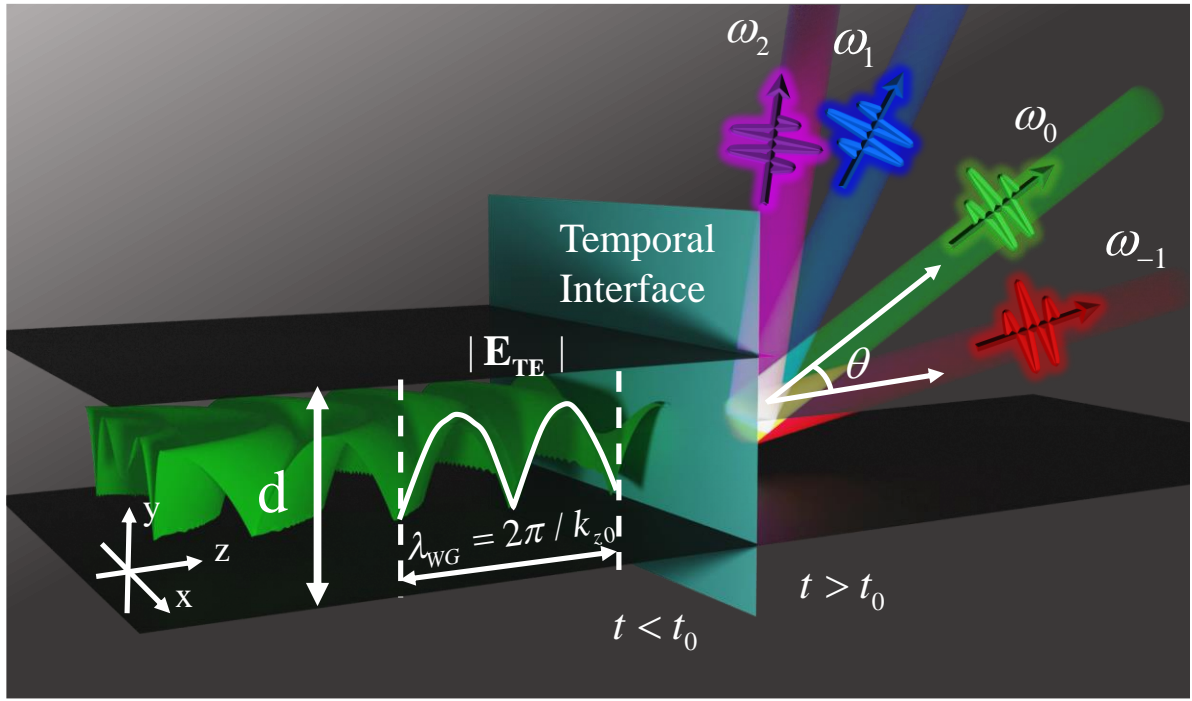

Fig.1: A parallel plate waveguide supporting a TE-mode is shown. At the temporal interface $t = t_0$, the top metallic plate changes its conductivity to zero, opening the waveguide towards the free space. A rainbow scattering process is triggered, the original monochromatic guided wave is converted into a radiated field characterized by different frequencies propagating along different directions.

In the last years, artificial engineered materials, namely metamaterials, have been used for realizing anomalous prims-like optical and electromagnetic responses by designing their frequency and spatial dispersive behaviors. For example, nonlocal metamaterials have been designed for inverting the relationship between the refractive index and illumination frequency, achieving an inverse prism with opposite functionality with respect to a canonical prism [2]; and space-time modulated metamaterials have been used for generating a discrete number of harmonics that are scattered in different directions [3]–[6]. In both cases, the control of the material properties in the spatial domain is fundamental for achieving the desired prism-like functionality. However, more recently, it has been demonstrated that the temporal control of the material properties is enough for achieving novel and interesting wave scattering phenomena, such as temporal aiming [7], temporal Fabry-Perot cavities [8], [9], anti-reflection temporal coatings [10]–[12], broadband absorbers [13], [14], and leaky-wave radiation [15], just to name a few. In this scenario, inverse prism based on a temporal metamaterial that switches instantaneously from an isotropic to an anisotropic medium has been proposed in [16], showing that the dispersion relation between temporal frequencies and spatial frequencies can be inverted. However, despite the spatial control is relaxed, the challenging implementation of an artificial media that switches its electromagnetic properties in the entire domain where the wave is propagating is still present. To overcome this issue, boundary-induced temporal interfaces have been proposed as a feasible implementation strategy for conceiving temporal metamaterials, where the temporal interface is realized without acting on the actual material properties, but rather on the effective ones [17], [18]. Indeed, quantities like the effective refractive index and effective wave impedance are typically used for describing the propagation characteristics within a guiding system, analogous to the wave propagation in an unbounded medium with the same effective material properties. In [18], the Authors demonstrated that a parallel plate waveguide can be effectively used for emulating a free-space temporal interface, realizing a frequency shift proportional to the jump of the effective refractive index perceived by the wave and a controlled scattering in forward and backward directions.

In this *Manuscript*, we extend the applications of the boundary-induced temporal interfaces proposing a temporal prism from whose interface a polychromatic rainbow-like radiation is scattered in free-space. We exploit the dispersive nature of a waveguide together with the unique properties of the temporal interfaces to realize a wavevector matching between a monochromatic guided wave and all the radiative modes at different frequencies that compose the polychromatic rainbow, as shown in the illustrative picture reported in Fig.1. Before the switching time $t_0$ (Fig. 2a), the guided mode is confined by two metallic parallel plates with ideal conductivity, *i.e.*, $\sigma \to \infty$, and propagates with a specific wavevector along the positive direction of the *z*-axis. At the switching time $t_0$ and following instants (Fig. 2b), the conductivity of the top plate goes to zero, making it electromagnetically transparent and exposing the guided mode to the free space. Since the switching is instantaneous, the spatial profile of the guided mode is conserved across the interface, realizing an effective temporal interface between two media: the first supports the propagation of modes according to the PPWG dispersion diagram and the second according to the free-space linear dispersion. The two sets of modes share the same projection of the free-space wavevector **k** onto the propagation direction $\hat{z}$ in both forward, $\omega > 0$, and backward directions, $\omega < 0$, with $\omega - k$ being the frequency of the scattered wave. In particular, Fig. 2b graphically illustrates how the rainbow is generated, showing that different wavenumbers *k*, *i.e.*, different frequencies being $\omega = kc$, are radiated towards different directions for satisfying the continuity of the *z*-component of the guided wavenumber across the interface.

In the following, we describe in closed form the relationship between the guided mode and the rainbow-like spectrum scattered from the boundary-induced temporal interface and derive the radiation direction in elevation and the scattering coefficients for each frequency component. Then, a design guideline for tailoring the temporal rainbow in terms of excited frequency components and propagation directions is reported and verified through a proper set of full-wave numerical simulations, confirming the design of the rainbow and the desired prism-like wave phenomenon.

## 2. Initial Conditions

Let us consider a PPWG formed by two metallic plates indefinitely extended in the *x*- and *z*-direction and placed at a distance $d$ in the *y*-direction. Hereafter we assume TE$_1$ propagation excited at frequency $n_b$ with wavelength $\lambda_{WG} = 2\pi / k_{z0}$ and propagation constant $k_{z0}$ given by the dispersion relation $k_{z0} = \sqrt{k_0^2 - k_y^2}$, where $k_0 = \omega_0 / c_0$ is the free-space wavenumber and $k_y = \pi / d$ is the transverse wavenumber [19]. In Fig. 2c the dispersion relation of the parallel plate waveguide for the TE$_1$ mode is graphically displayed. By fixing the transverse wavenumber $k_y = \pi / d$, the light-cone is cut in the $k_z\omega$-plane identifying two hyperbolae for positive and negative frequencies, respectively, that describe the dispersion relation for the guided modes. Being the excitation frequency always a positive quantity, from the upper hyperbola we can easily derive the wavenumbers $\pm k_{z0}$ at the excitation frequency $\omega_0$ for the supported forward and backward propagating guided waves, respectively. At the instant of time $t = t_0$, the conductivity of the top metallic plate changes instantaneously to zero, realizing a temporal interface for the guided mode and triggering the desired rainbow temporal scattering.

### 2.1. Rainbow Scattered Frequencies

To analyze the rainbow scattering process, the continuity of the induction fields **B**, **D** and the preservation of the wavelength [20]–[23] across the interface must be imposed. Also in this case, we can describe the underlying physics through the light-cones in Fig. 2. Preserving the wavelength results in preserving the propagation constant $k_{z0}$, leading to find the eigensolutions for the radiated field at the intersection between the light-cone and a $k_y\omega$-plane (with $k_z = k_{z0}$) obtaining another pair of hyperbolae, as shown in Fig. 2d. After the temporal interface, the guided wave acts as a spatially distributed source for the radiated field and, therefore, the two hyperbolae in Fig. 2d are now describing all possible frequencies $\omega$, and the corresponding transverse wavenumbers $k_y$, that the distributed source can excite simultaneously. Therefore, we can derive the frequency of the plane waves radiated after the temporal interface towards the elevation angle $\theta$, defined with respect to the positive direction of the *z*-axis. By imposing conservation of the spatial frequency across the interface, the projection of the wavevector **k** of the generic radiated plane wave on the *z*-axis is always:

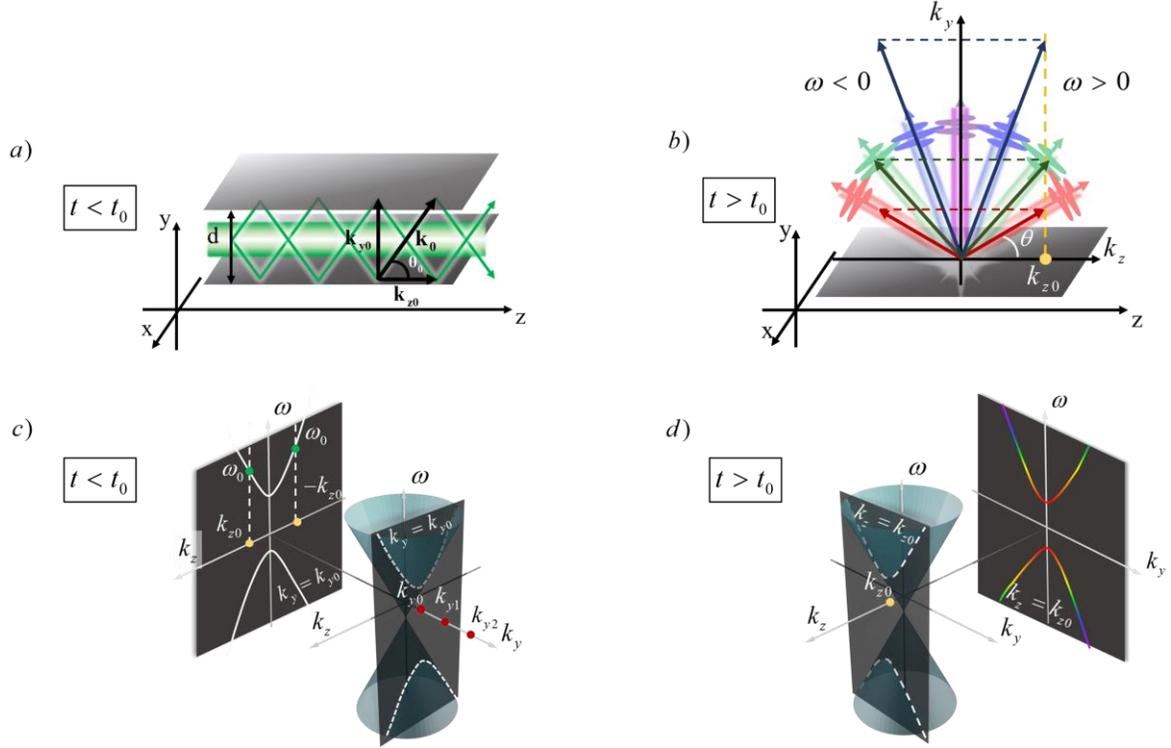

Fig.2: a parallel plate waveguide (PPWG) of height $d_1$ is fed with frequency $\omega_0$ at $t < t_0$ in (a). The field distribution can be seen as the interference between two waves propagating along the z-direction. After the temporal interface, $t > t_0$, the upper metallic plate changes its conductivity from $\sigma = +\infty$ to $\sigma = 0$ triggering the rainbow temporal scattering in both forward $\omega > 0$ and backward $\omega < 0$ direction, (b). The coupling between the starting field and the free space radiation is dictated by the projection of the $\mathbf{k}$ vector onto the propagation direction z, its projection must be equal to starting propagation constant $k_{z0}$. The dispersion relation of the PPWG can be seen as a cut on the light cone in the plane $\omega k_z$, e.g., constant $k_y$, (c). Vice versa, the temporal interface preserves the propagation constant $k_{z0}$ which means the cut on the light cone is on the $\omega k_y$ plane, (d).

$$\mathbf{k} \circ \hat{z} = k \cos\theta = k_{z0}, \qquad (1)$$

where $\circ$ indicates the dot product. Making use of the free-space dispersion relation $k = \omega / c_0$ and (1) we find the relation $\omega(\theta)$ that describes the frequency content of the temporal scattered rainbow for each angle from the original propagation constant of the guided mode:

$$\omega(\theta) = \frac{k_{z0} c_0}{\cos(\theta)}. \qquad (2)$$

Equation (2) returns also the frequency range of our rainbow from the minimum frequency $\omega_{min} = k_{z0} c_0$ of the wave propagating along the metallic plane (*i.e.*, $\theta = 0°$) to theoretically the maximum frequency $\omega_{max} \to +\infty$ for the wave propagating in broadside direction (*i.e.*, $\theta = 90°$).

## 2.2. Rainbow Scattering Parameters

Let us now focus our attention on the wavevector matching between the original guided mode and the different plane waves composing the rainbow. Despite eq. (2) suggests that plane waves at extremely high frequencies can be excited, we have to consider that very few energy can be coupled in that direction due to the high wavevector mismatching. To quantify the amount of energy coupled to each radiated plane wave, we must impose the continuity of the induction field across the interface. In our scenario, assuming propagation in vacuum both in the PPWG and in free space, the boundary conditions for the induction fields reduce to the ones for the **H**, **E** fields as:

$$H_y(y,z)^{t_0^-} = H_y(y,z)^{t_0^+}$$
$$H_z(y,z)^{t_0^-} = H_z(y,z)^{t_0^+}, \quad (3)$$
$$E_x(y,z)^{t_0^-} = E_x(y,z)^{t_0^+}$$

where $[H_y, H_z, E_x]$ is the TE$_1$ triplet inside the PPWG and $t_0^-$ and $t_0^+$ are the instants of time right before and right after the temporal interface at $t = t_0$, respectively. The starting field inside the waveguide can be seen as the interference pattern between two waves, one with positive transverse wavenumber $+k_{y0}$ and the other with negative transverse wavenumber $-k_{y0}$, as shown in Fig. 2a. On the other hand, the free space field can be given by the superposition of two continuous sets of plane waves, one set travelling with positive frequencies $\omega > 0$, *i.e.*, forward direction, and the other set travelling with negative frequencies $\omega < 0$, *i.e.*, backward direction, but all along the positive *y*-direction, as shown in Fig. 2b. Actually, we should also consider the two sets of plane waves for the propagation along the negative *y*-direction, but the presence of the bottom metallic plate limits the domain to the upper half space. Moreover, the metallic plate reflects the wave characterized by $-k_{y0}$. The distribution arising from their interference pattern is the difference between two complex exponentials with opposite arguments, hence a sine function of *y*-coordinate. We can expand in plane waves the free-space fields as follows:

$$H_y(y,z)^{t_0^+} = \cos(\theta)\left(\int_0^{+\infty} \text{FW}(k_y)\sin(k_y y)e^{-ik_{z0}z}e^{i\omega^+(t-t_0)}dk_y\right.$$
$$\left.-\int_0^{+\infty} \text{BW}(k_y)\sin(k_y y)e^{-ik_{z0}z}e^{-i\omega^-(t-t_0)}dk_y\right)$$
$$E_x(y,z)^{t_0^+} = \int_0^{+\infty} \text{FW}(k_y)\sin(k_y y)e^{-ik_{z0}z}e^{i\omega^+(t-t_0)}dk_y$$
$$+\int_0^{+\infty} \text{BW}(k_y)\sin(k_y y)e^{-ik_{z0}z}e^{-i\omega^-(t-t_0)}dk_y \qquad (4)$$

where $\cos(\theta)$ is the normalized TE wave impedance and $[\text{FW},\omega^+]$, $[\text{BW},\omega^-]$ stand for the amplitude and frequency of the forward scattered and backward scattered waves, respectively. Here we omitted the magnetic field along the z-direction $H_z$, which depends on $[H_y, E_x]$ in (4). Using the plane wave decomposition, the final scattered parameters can be written as (please, refer to the Appendix section for details):

$$\text{FW} = \frac{1}{2k_{y0}}\left(\frac{\omega_0}{c_0} + \frac{k_z}{\cos(\theta)}\right)\kappa(\theta,k_y)$$
$$\text{BW} = \frac{1}{2k_{y0}}\left(\frac{\omega_0}{c_0} - \frac{k_z}{\cos(\theta)}\right)\kappa(\theta,k_y) \qquad (5)$$

where $\omega_0$ is given by the overlap integral between the initial distribution and the corresponding plane wave in the $n_a$ direction. An interesting observation can be made on (5): in fact, the maximum of the forward scattering coefficient is in the direction identified by the starting wavevector $\mathbf{k}_0$ inside the PPWG shown in Fig. 2a; moreover, equation (2) reveals that the plane wave radiated towards this direction is the original frequency $\omega_0$ of the guided mode.

## 3. Rainbow Design

In this section, we derive an effective design guideline using the analytical model presented in the previous section. Let us start from equation (5) that reveals the direction of maximum radiation for the forward wave that corresponds to the angle $\theta_0$ between the free-space wavevector inside the PPWG and the propagation direction (Fig. 2a). From the guided wave theory [24], we can write:

$$\tan(\theta_0) = \frac{1}{\sqrt{(\omega_0/\omega_{cut})^2 - 1}},\qquad(6)$$

$$\omega|_{\theta=\theta_0} = \omega_0$$

where $\omega_{cut}$ is the cut-off frequency of the propagating TE$_1$ mode within the PPWG dictated by the transverse dimension $d$. Therefore, the transverse dimension $d$ can be used for designing the principal direction of our rainbow, being $d$ related to the transverse wavevector and, thus, for a given frequency $\omega_0$, the corresponding elevation angle $\theta_0$. After some algebraic manipulation, we can relate the generic dimension $d$ of the PPWG to the elevation angle $\theta$ where the plane-wave at frequency $\omega_0$ must be scattered as follows:

$$d = \frac{\pi c_0}{\omega_0 \sin(\theta)} = \frac{\lambda_0}{2\sin(\theta)},\qquad(7)$$

where $\lambda_0 = c_0/f_0$ is the free-space wavelength at frequency $f_0 = \omega_0/2\pi$. Equation (7) can now be used for designing the guiding system and achieving the desired distribution of the scattered rainbow. However, it is worth highlighting that a monomodal regime must be always satisfied in the original PPWG, which implies that the exciting frequency $\omega_0$ cannot exceed the cutoff frequency of the second higher order mode. Under this constrain, the angular range within the main plane wave can be radiated is $[30°, 90°]$.

To test the design guideline, the plane wave component of our rainbow after the temporal interface at $f_0 = 10\,\text{GHz}$ must be radiated towards $\theta = 50°$. Applying (7), we should set the dimension of the PPWG to $d = 20\,\text{mm}$. For such a waveguide, the cut-off frequency of the higher order mode is $f = 15\,\text{GHz}$, hence higher than the exciting frequency, satisfying the monomodal regime.

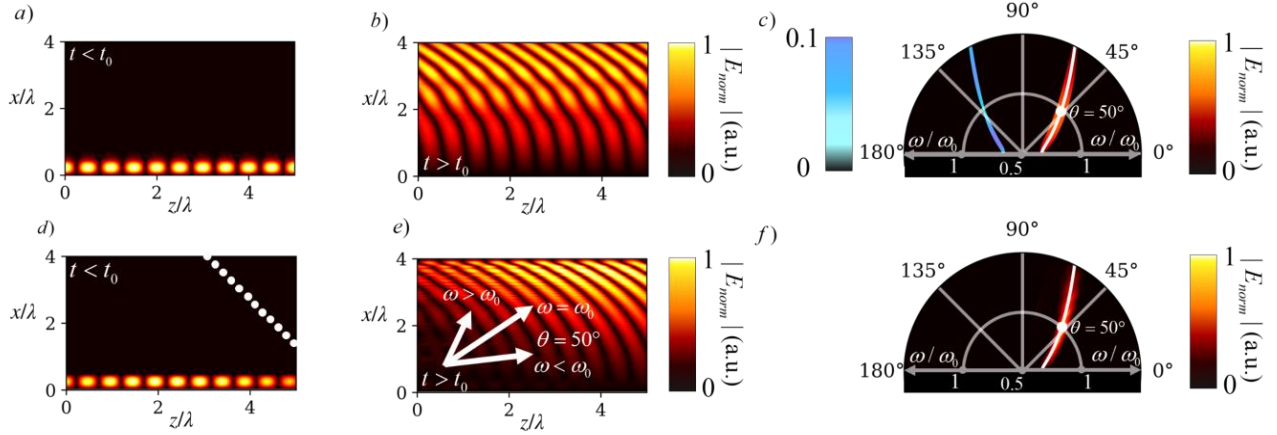

Fig. 3, the starting field inside the PPWG is reconstructed for $t < t_0$, (a), and then scattered for $t > t_0$, (b), with the use of the analytically found scattering coefficients. The spectrum of the temporal rainbow is reported in polar coordinates using the normalized frequency $\omega/\omega_0$ as radius and $\theta$ as angular variable. The principal direction of scattering is at the design direction $\theta = 50°$ and at the expected normalized frequency $\omega/\omega_0 = 1$, (c). The backward scattered rainbow, in the interval $[90°, 180°]$, is loosely coupled with the starting field and had to be plotted with a different colormap, in blue, and different scale, an order of magnitude lower than the forward rainbow. In (d)-(f), the same results are reported extracted from the numerical simulation. The designed rainbow is confirmed also by the FDTD simulations.

## 4. Numerical Experiments

In this section, we demonstrate the validity of the analytical model describing the scattering at a boundary-induced temporal interface and the raising of the expected rainbow-like scattering by using a proper set of finite difference time domain (FDTD) full-wave numerical simulations [25].

In Fig. 3a-c, we report the analytical results for the PPWG with dimension $d = 20$ mm and source $f = 10$ GHz. In particular, in Fig. 3a and 3b we report the reconstructed fields inside the waveguide $(t < t_0)$ and in the free-space, $(t > t_0)$ using (4) and (5). In particular, Fig. 3a shows perfect reconstruction of the field inside the waveguide for $t < t_0$, while Fig. 3b shows the scattered rainbow at $t > t_0$, where higher frequencies are clearly visible for higher elevation angles. Finally, in Fig. 3c, we report the polar density plot where the amplitude of the scattered waves as given by (5) are shown as a function of the normalized frequency $\omega/\omega_0$ (radial dimension of the plot) and the angle $\theta$ (angular dimension of the plot). It can be noticed that the amplitude of the backward scattered waves is very low with respect to the forward waves due to the lower coupling with the original guided field in the waveguide. Moreover, Fig. 3c confirms that the principal direction of scattering occurs at $\omega/\omega_0 = 1$ and towards $\theta = 50°$, as derived in the design section.

Figs. 3d-f report the results of the full-wave numerical experiments. In the FDTD simulation domain, the metallic plates of the PPWG are implemented as PEC boundary conditions and a linear array of probes is placed in free-space to sample the plane wave contribution of the rainbow propagating towards its normal direction. In Figs. 3d-e we report two snapshots in time of the field inside the waveguide and the scattered field, respectively. The agreement between the numerically computed fields, Fig. 3d-e, and the analytically reconstructed fields is almost perfect. Finally, a polar density plot where the amplitude of the scattered waves is also derived from the numerical simulation is reported in Fig. 3f. Also in this case, the principal direction of the rainbow scattering is towards $\theta = 50°$, confirming the design formulas.

## 5. Conclusions

In this *Manuscript*, we have reported a novel wave phenomenon based on temporal interfaces that gives rise to a rainbow temporal scattering into free-space. After having introduced and detailed the physics behind this phenomenon, we have analytically modeled it and derived a design guideline for controlling the main scattering direction of the rainbow. In the context of temporal metamaterials, this Manuscript provides closed form solutions for scenarios that have not been explored yet. Moreover, this implementation will ease the requirements in realizing devices based on temporal metamaterials thanks to the fact that we work only with a surface boundary rather than with the bulk material.

# APPENDIX

In this appendix we report the derivation of the scattering parameters reported in (5). At the temporal interface, we must impose the continuity of the induction fields $\mathbf{B}, \mathbf{D}$, which, in our scenario, relaxes to the $\mathbf{H}, \mathbf{E}$ fields, as mentioned earlier. In other words, we must solve the set of equations reported in (3). From [19], the TE field inside the waveguide is described by:

$$H_y(y,z)^{t_0^-} = i\frac{\omega_0}{k_{y0}c_0}\sin(k_{y0}y)e^{-ik_{z0}z}e^{i\omega_0 t}$$
$$H_x(y,z)^{t_0^-} = \cos(k_{y0}y)e^{-ik_{z0}z}e^{i\omega_0 t} \quad (A1)$$
$$E_x(y,z)^{t_0^-} = i\frac{k_{z0}}{k_{y0}}\sin(k_{y0}y)e^{-ik_{z0}z}e^{i\omega_0 t}$$

Inserting (4) and (A1) in $H_y$ and $E_x$ in (3), we get the following system of equations:

$$i\frac{\omega_0}{k_{y0}c_0}\sin(k_{y0}y)e^{-ik_{z0}z}e^{i\omega_0 t} = \cos(\theta)\left(\int_0^{+\infty}\text{FW}(k_y)\sin(k_y y)e^{-ik_{z0}z}e^{i\omega^+(t-t_0)}dk_y\right.$$
$$\left. - \int_0^{+\infty}\text{BW}(k_y)\sin(k_y y)e^{-ik_{z0}z}e^{-i\omega^-(t-t_0)}dk_y\right) \quad (A2)$$
$$i\frac{k_{z0}}{k_{y0}}\sin(k_{y0}y)e^{-ik_{z0}z}e^{i\omega_0 t} = \int_0^{+\infty}\text{FW}(k_y)\sin(k_y y)e^{-ik_{z0}z}e^{i\omega^+(t-t_0)}dk_y$$
$$+ \int_0^{+\infty}\text{BW}(k_y)\sin(k_y y)e^{-ik_{z0}z}e^{-i\omega^-(t-t_0)}dk_y$$

After some algebraic manipulations, we can isolate the integral terms containing the forward and backward scattering coefficients:

$$\int_0^{+\infty}\text{FW}(k_y)\sin(k_y y)e^{-ik_{z0}z}e^{i\omega^+(t-t_0)}dk_y =$$
$$= \frac{i}{2k_{y0}}\left(\frac{\omega_0}{c_0} + \frac{k_{z0}}{\cos(\theta)}\right)\sin(k_{y0}y)e^{-ik_{z0}z}e^{i\omega_0 t}$$
$$\int_0^{+\infty}\text{BW}(k_y)\sin(k_y y)e^{-ik_{z0}z}e^{-i\omega^-(t-t_0)}dk_y = \quad (A3)$$
$$= \frac{i}{2k_{y0}}\left(\frac{\omega_0}{c_0} - \frac{k_{z0}}{\cos(\theta)}\right)\sin(k_{y0}y)e^{-ik_{z0}z}e^{i\omega_0 t}$$

At this point, it is possible to invert the plane wave decomposition from the left side, where we integrate over all possible $k_y$ to the right side, where we integrate over the section $d$ together with dropping the spatial dependency from z and the time dependency from t:

$$\text{FW}(k_y) = \frac{i}{2k_{y0}}\left(\frac{\omega_0}{c_0} + \frac{k_{z0}}{\cos(\theta)}\right)\int_0^d \sin(k_{y0}y)\sin(k_y y)dy$$
$$\text{BW}(k_y) = \frac{i}{2k_{y0}}\left(\frac{\omega_0}{c_0} - \frac{k_{z0}}{\cos(\theta)}\right)\int_0^d \sin(k_{y0}y)\sin(k_y y)dy \quad (A4)$$

The integral over the section on the right-hand side can be solved in closed form as:

$$\kappa(\theta, k_y) = \frac{\sin\left((k_{y0} - k_y)d\right)}{2(k_{y0} - k_y)} - \frac{\sin\left((k_{y0} + k_y)d\right)}{2(k_{y0} + k_y)}, \quad (A5)$$

which has a maximum for $k_y = k_{y0}$, hence, the principal direction of scattering coincides with the angle of the starting waves propagating inside the PPWG. To conclude, the final expressions for the scattering coefficients are:

$$\begin{aligned} \text{FW} &= \frac{1}{2k_{y0}} \left( \frac{\omega_0}{c_0} + \frac{k_z}{\cos(\theta)} \right) \kappa(\theta, k_y) \\ \text{BW} &= \frac{1}{2k_{y0}} \left( \frac{\omega_0}{c_0} - \frac{k_z}{\cos(\theta)} \right) \kappa(\theta, k_y) \end{aligned} \quad (A6)$$